 \documentclass[final,3p,times,number]{elsarticle}

\usepackage{amssymb}
\usepackage{soul}
\usepackage{lipsum}
\usepackage{mathtools,leftindex,tensor,mhchem}
\usepackage{graphicx}
\usepackage{subcaption}
\usepackage{CJKutf8}
\usepackage{float}

\usepackage{amsthm}
\usepackage{hyperref}
\usepackage{soul}

\usepackage{xcolor}

\usepackage{etoolbox}
\makeatletter
\def\ps@pprintTitle{%
  \let\@oddhead\@empty
  \let\@evenhead\@empty
  \let\@oddfoot\@empty
  \let\@evenfoot\@oddfoot
}
\makeatother

\journal{Nuclear Physics A}

\begin{document}
\setcitestyle{square}
\setcitestyle{comma}
\begin{frontmatter}


\title{Investigation of the Excited States of \ce{^{114}Sn} Using the GRIFFIN Spectrometer at TRIUMF
}

%% \affiliation[label1]{organization={},The semi-magic 110 – 122Sn isotopes display signs of shape coexistence, in their excited states that are close in energy to the spherical
% ground states. This paper investigates the nuclear structure of 114Sn using the competing β+-decay and electron capture of a
% radioactive beam of 114Sb produced at the TRIUMF-ISAC facility located in Vancouver, at which the Gamma-Ray Infrastructure
% For Fundamental Investigations of Nuclei (GRIFFIN) spectrometer was utilized. The investigation will allow for an in-depth
% understanding of the excited 0+ states in 114Sn and their decay patterns, and is part of a campaign to study shape coexistence in the
% Sn isotopes using β-decay.
\author[SFUp]{N. K. Syeda}
\author[SFU]{P. Spagnoletti}
\author[SFU]{C. Andreoiu}
\author[CSNSM]{C.M. Petrache}
\author[SFU]{D. Annen}
\author[TRIUMF]{R.S. Lubna}
\author[TRIUMF]{V. Vedia}
\author[IFIC]{A. Algora}
\author[wl]{A. Babu}
\author[TRIUMF]{G.C. Ball}
\author[TRIUMF]{S. Bhattacharjee}
\author[TRIUMF]{R. Caballero-Folch}
\author[U of Guelph]{R. Coleman}
\author[TRIUMF,Victoriaaa]{I. Dillmann}
\author[Regina]{E.G. Fuakye}
\author[U of Liverpool]{L.P. Gaffney}
\author[SFU]{F. H. Garcia}
\author[TRIUMF]{A.B. Garnsworthy}
\author[U of Guelph]{P.E. Garrett}
\author[TRIUMF]{C.J. Griffin}
\author[Regina]{G.F. Grinyer}
\author[TRIUMF]{G. Hackman}
\author[SMU]{R. Kanungo}
\author[Regina]{K. Kapoor}
\author[U of Guelph]{A. Laffoley}
\author[TRIUMF]{G. Leckenby}
\author[U of Guelph]{K. Mashtakov}
\author[TRIUMF, Mines]{C. Natzke}
\author[TRIUMF]{B. Olaizola}
\author[SFU]{K. Ortner}
\author[TRIUMF]{C. Porzio}
\author[U of Guelph]{M. Rocchini}
\author[Regina]{N. Saei}
\author[TRIUMF]{Y. Saito}
\author[U of Liverpool]{M. Satrazani}
\author[Regina]{D. Shah}
\author[ANL]{M. Siciliano}
\author[japan]{J. Smallcombe}
\author[TRIUMF,U of Guelph]{C.E. Svensson}
\author[Regina]{A. Talebitaher}
\author[TRIUMF]{R. Umashankar}
\author[U of Guelph]{S. Valbuena Burbano}
\author[TRIUMF]{J. Williams}
\author[SFU]{\begin{CJK}{UTF8}{gbsn}
F. Wu (吴桐安)
\end{CJK}}
\author[TRIUMF]{D. Yates}
\author[U of Guelph]{T. Zidar}
% \author[U Tohoku,TRIUMF]{R. Kono}
% \author[IIT]{S. Das}
% \author[wl]{M. Berube}
% \author[Grenoble INP - Phelma]{N. Tanzi}
% \author[TRIUMF]{J. R. Murias}University of Regina

\affiliation[SFUp]{organization={Department of Physics},%Department and Organization
            addressline={Simon Fraser University}, 
            city={Burnaby},
            postcode={V5A 1S6}, 
            state={BC},
            country={Canada}}
            
\affiliation[SFU]{organization={Department of Chemistry},%Department and Organization
            addressline={Simon Fraser University}, 
            city={Burnaby},
            postcode={V5A 1S6}, 
            state={BC},
            country={Canada}}

\affiliation[CSNSM]{organization={Université Paris-Saclay, CNRS/IN2P3, IJCLab},%Department and Organization
            % addressline={4004 Wesbrook Mall}, 
            city={Orsay},
            % postcode={V6T 2A3}, 
            % state={BC},
            country={France}}

\affiliation[TRIUMF]{organization={TRIUMF},%Department and Organization
            addressline={4004 Wesbrook Mall}, 
            city={Vancouver},
            postcode={V6T 2A3}, 
            state={BC},
            country={Canada}}

\affiliation[IFIC]{organization={Instituto de Física Corpuscular},%Department and Organization
            addressline={Carrer del Catedrátic José Beltrán Martinez, 2}, 
            city={46980 València},
            % postcode={N1G 2W1}, 
            % state={Ontario},
            country={Spain}}

\affiliation[wl]{organization={Department of Physics},%Department and Organization
            addressline={University of Waterloo}, 
            city={Waterloo},
            postcode={N2L 3G1}, 
            state={Ontario},
            country={Canada}}

\affiliation[U of Guelph]{organization={Department of Physics},%Department and Organization
            addressline={University of Guelph}, 
            city={Guelph},
            postcode={N1G 2W1}, 
            state={Ontario},
            country={Canada}}

\affiliation[Victoriaaa]{organization={Department of Physics},%Department and Organization
            addressline={University of Victoria}, 
            city={Victoria},
            postcode={V8P 5C2}, 
            state={British Columbia},
            country={Canada}}

\affiliation[Regina]{organization={Department of Physics},%Department and Organization
            addressline={University of Regina}, 
            city={Regina},
            postcode={S4S 0A2}, 
            state={Saskatchewan},
            country={Canada}}

\affiliation[U of Liverpool]{organization={University of Liverpool},%Department and Organization
            addressline={Brownlow Hill}, 
            city={Liverpool L69 7ZX},
            % postcode={N1G 2W1}, 
            % state={Ontario},
            country={United Kingdom}}

\affiliation[SMU]{organization={Department of Physics},%Department and Organization
            addressline={Saint Mary's University}, 
            city={Halifax},
            postcode={B3H 3C3}, 
            state={Nova Scotia},
            country={Canada}}

\affiliation[Mines]{organization={Department of Physics},%Department and Organization
            addressline={Colorado School of Mines}, 
            city={Golden},
            postcode={80401}, 
            state={Colorado},
            country={USA}}

\affiliation[ANL]{organization={Physics Division},%Department and Organization
            addressline={Argonne National Laboratory}, 
            city={Lemont},
            postcode={60439}, 
            state={Illinois},
            country={USA}}

\affiliation[japan]{organization={Advanced Science Research Center, Japan Atomic Energy Agency, Tokai, Ibaraki 319-1195 },%Department and Organization
            % addressline={Simon Fraser University}, 
            % city={Burnaby},
            % postcode={V5A 1S6}, 
            % state={BC},
            country={Japan}}
% \affiliation[IIT]{organization={Indian Institute of Technology Indore
% },%Department and Organization
%             % addressline={Indian Institute of Technology Indore}, 
%             city={Indore},
%             state={Madhya Pradesh},
%             country={India}}

\begin{abstract}
The semi-magic \ce{^{110-122}Sn} isotopes display signs of shape coexistence in their excited $0^+$ states, which, in contrast to the spherical $0^+$ ground states, are deformed.
%that are close in energy to the spherical ground $0^+$ 
This paper investigates the nuclear structure of \ce{^{114}Sn} using the competing $\beta^{+}$ decay and electron capture of a radioactive beam of \ce{^{114}Sb} produced at the TRIUMF-ISAC facility using the GRIFFIN spectrometer. This study will allow for an in-depth understanding of the excited $0^{+}$ states in \ce{^{114}Sn}, by focusing on 
% the excited $0^{+}$  states and
their decay patterns.
%will provide insights into shape coexistence and intruder configurations in \ce{^{114}Sn}.
In the present experiment, transitions at 856.2-keV and 1405.0-keV, which were observed in an earlier $\beta^{+}$ decay study but not placed in the \ce{^{114}Sn} level scheme, have been assigned to the level scheme in connection to the $0_3^+$ level at 2156.0-keV. Properly assigning these transitions refines the level scheme and enhances our understanding of the nuclear structure in \ce{^{114}Sn}.

%5and is part of a campaign to study shape coexistence in the Sn isotopes using $\beta$-decay.
\end{abstract}

\end{frontmatter}
\section{Introduction}\label{introducton}

%\underline{Excites states in $^{114}$Sn were populated by the $\beta^{+}$ decay of $\^{114}$Sb at the TRIUMF-ISOL facility and the subsequent $\gamma$-decay was observed using the GRIFFIN spectrometer.}

With the largest number of stable isotopes of any element across the nuclear chart and spanning beyond two full neutron shell closures, the tin isotopic chain is an ideal testing ground for theoretical models. The semi-magic Sn isotopes, are of significant interest as they serve as benchmark nuclei for state-of-the-art shell-model calculations, and provide a solid foundation for understanding shape evolution in the \textit{Z} = 50 region. Although these isotopes are nearly spherical in their ground state, they exhibit deformed bands built on excited $0^{+}$ states, which are observed throughout the chain in the neutron mid-shell and can be characterized by two-particle-two-hole (\textit{2p-2h}) configurations. The presence of these deformed intruder states is a key feature in the \textit{Z} = 50 region, and further investigation is needed to explore the degree of mixing between deformed and normal states \cite{Nuclears3:online}. Situated at the neutron mid-shell between \textit{N} = 50 and \textit{N} = 82, \ce{^{114}Sn}, is a prime candidate for detailed spectroscopic studies of nuclear shape coexistence \cite{Pore2017}. 
%\underline{Excites states in $^{114}$Sn were populated by the $\beta^{+}$ decay of $\^{114}$Sb at the TRIUMF-ISOL facility and the subsequent $\gamma$-decay was observed using the GRIFFIN spectrometer.}

To investigate the decay properties of the \textit{2p-2h} band in \ce{^{114}Sn}, a high-statistics experiment was conducted, populating excited states in \ce{^{114}Sn} through the $\beta^+$\slash electron capture decay of the \ce{^{114}Sb} parent to study energy levels and $\gamma$-ray transitions.

\section{GRIFFIN Experiment}
\begin{figure}[h]
	\centering 
	\includegraphics[width=0.48\textwidth]{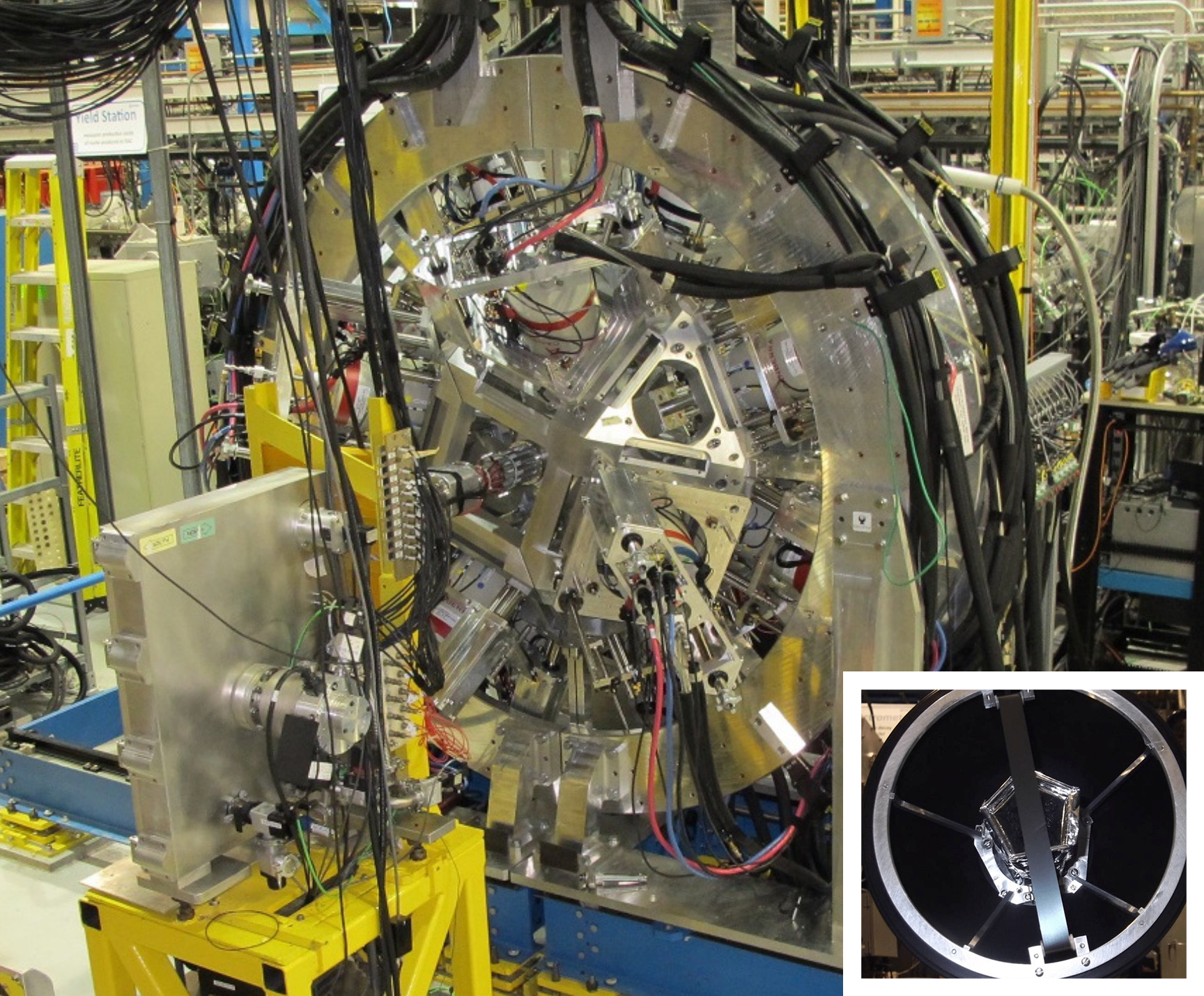}	
	\caption{The GRIFFIN spectrometer located in the ISAC-I hall at TRIUMF. The radioactive \ce{^{114}Sb} beam was implanted in the Mylar tape at the center of the decay chamber shown here in the lower right inset. The tape is moving according to a pre-programmed duty cycle to create a new implantation spot, and then stored in the lead box seen at the back of GRIFFIN to shield the HPGe detectors from the activity built on the tape. Image is adapted from Ref. \cite{griffin2020} .} 
	\label{griffin}%
\end{figure}

%%\label{}
The experiment was conducted using the Gamma-Ray Infrastructure For Fundamental Investigations of Nuclei (GRIFFIN) spectrometer, shown in Fig.~\ref{griffin}, at the TRIUMF-ISAC facility in Vancouver, Canada \cite{GARNSWORTHY20199}. GRIFFIN is a state-of-the-art $\gamma$-ray spectrometer, which contains up to 16 Compton-suppressed large-volume HPGe clover detectors for $\gamma$-ray measurements, arranged in a nearly $4\pi$ configuration around an implantation chamber, ensuring high efficiency and angular coverage \cite{rizwan}. Each of the 16 clover detectors that make up GRIFFIN is composed of four n-type HPGe crystals, each measuring 60 mm in diameter and 90 mm in length \cite{rizwan}. 

 A radioactive beam of \ce{^{114}Sb} with a half-life $(t_{1/2})$ of 3.49~min was produced using the Isotope Separator On-Line (ISOL) technique where 10 µA protons at 480 MeV were directed onto a uranium carbide (UCx) target. The TRIUMF Resonant Ionization Laser Ion Source (TRILIS) was used to selectively ionize the species of interest \cite{LASSEN2023137}. The secondary beam was then directed to the mass separator, to select \textit{A} = 114 ions. 
 
% purify the beam by suppressing isobaric contaminants Ref \cite{LASSEN2023137}.

The high-resolution mass separator, utilizing a constant magnetic field, guided the ions through a 45-degree bend. The trajectory of the ions was determined by their mass-to-charge ratio. Adjustable slits at the end of the separator allowed for the selection of ions of a specific mass with a resolving power of $M/ \Delta M \approx 2000$~\cite{Bricault2014}. 

In this experiment, ions with a mass number of \textit{A} = 114 were selected resulting in a radioactive beam that included \ce{^{114}Sb} (I$^\pi$=3$^{+}$, $t_{1/2} = 3.49$ m)~\cite{jean},  \ce{^{114}In}$^{m1}$ (I$^\pi$=5$^{+}$, ${t_{1/2}}$ = 49.5 days)~\cite{jean},  \ce{^{114}In}$^{m2}$ (I$^\pi$=8$^{-}$, ${t_{1/2}}$ = 43.1 ms))~\cite{jean}, \ce{^{114}In} (I$^\pi$=1$^{+}$, $t_{1/2}= 71.9$ s)~\cite{jean}, and \ce{^{95}Sr^{19}F} (I$^\pi$=3/2$^+$ $t_{1/2}= 23.9 $ s)~\cite{jean1} molecules.

This cocktail beam was then transported and implanted into Mylar tape at the center of the GRIFFIN chamber, shown in Fig.~\ref{griffin}, at a rate of $5 \times 10^5$ pps.  The decay cycle was set to 390 seconds of implantation followed by 390 seconds of decay, optimizing the detection of the $3^+$ state in \ce{^{114}Sb} that populates the states of interest in \ce{^{114}Sn}. After each decay cycle, the tape was moved into a lead box, at the back of the GRIFFIN Spectrometer. 

The energy and efficiency calibration of GRIFFIN was performed using \ce{^{56}Co}, \ce{^{60}Co}, \ce{^{133}Ba} and \ce{^{152}Eu} standard sources, with energies spanning from 81 keV to 3.6 MeV. Summing corrections as discussed in Ref. \cite{GARNSWORTHY20199} were also performed. The GRIFFIN efficiency for detecting $\gamma$-rays in add-back mode was 9$\%$ at 1 MeV. Only 15 HPGe clovers
were used in the present work.
The experiment was conducted over a period of 48 hours, during which a total of $2 \times 10^{10}$ $\gamma$-rays  were recorded with the beam on, and $8 \times 10^{9}$ events were captured during beam-off periods. A 250 ns coincidence window was applied to these data to construct the level scheme.

%The experiment ran for 48 hours, during which $2\times 10^{10}$ $\gamma$-ray single events were recorded and a 250 ns coincidence window was used to build the level scheme.

\section{Preliminary Results}
The last $\beta^{+}-$decay study of $\ce{^{114}Sn}$ was published in 1976 by Wigmans \textit{et al}. \cite{PhysRevC.14.229}. Even though the excited $\ce{0_{2}^{+}}$ 1953.1-keV level was observed to decay through a transition at 653.3-keV, the excited $\ce{0_{3}^{+}}$ 2156.0-keV level was not observed at that time. The 2156.0-keV level was later observed in a Coulomb excitation experiment~\cite{osti_4841795}, decaying by a 856.2-keV transition to the $\ce{2_1^+}$ 1299.8-keV level. More recently, the level scheme of $\ce{^{114}Sn}$ has been extended up to 4022.4-keV in a $\ce{^{114}Sn}(p,p^{'}\gamma)$ experiment, in which the phenomenon of shape coexistence was also investigated by measuring the nuclear lifetimes using the Doppler-shift attenuation coincidence technique \cite{PhysRevC.97.054319}. Building upon these findings, this experiment aims to expand the level scheme of $\ce{^{114}Sn}$ beyond the 4022.4-keV level using the GRIFFIN spectrometer.
%Building upon these findings and aiming to further investigate the nuclear structure of $\ce{^{114}Sn}$, particularly at higher excitation energies, using the GRIFFIN spectrometer, the level scheme is to be expanded further.

%to extract reduced transition probabilities. This work focuses on the decay pattern of the levels built on the excited $\ce{0_{2}^{+}}$ $1953.27~$-keV state in a selective $\beta^{+}$-decay experiment, utilizing a high-resolution spectrometer to extract comprehensive spectroscopic information.

\begin{figure}[H]
	\centering 
	\includegraphics[width=1\textwidth]{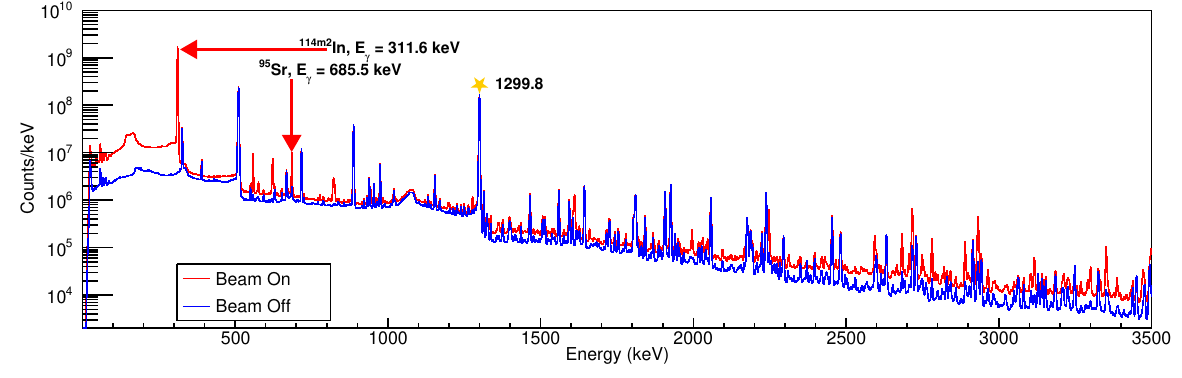}
\caption{Gamma-ray singles spectra with add-back for beam-on and beam-off. The yellow star indicates the $2^+_1 \rightarrow 0_1^+$ 1299.8-keV transition in $\ce{^{114}Sn}$. The 311.6-keV $\gamma$-ray seen in the beam-on spectrum is associated with the $8^-\rightarrow5^+$ isomeric transition in $\ce{^{114}In}$ and disappears in the beam-off spectrum. The transition at 685.5-keV is associated with the $\beta^-$-decay of $\ce{^{95}Sr}$ to $\ce{^{95}Y}$.}
	\label{beam}%
\end{figure}

Figure \ref{beam} shows overlapped $\gamma$-ray singles spectra with add-back for beam-on (red) and beam-off (blue). The data  demonstrates a significant reduction in background when the beam is off. Notably, the 311.6-keV $\gamma$-ray transition associated with the decay from the 8$^-$ to 5$^+$ isomeric state in $\ce{^{114}In}$ is absent in the beam-off spectrum, due to the short half-life nature of the 8$^-$ isomeric state. Additionally, the 685.5-keV peak, associated with the $\ce{^{95}Sr}$  $\beta^--$decay to $\ce{^{95}Y}$, is highly suppressed once the beam is off, further showing the high selectivity of our experiment. 

\begin{figure*}[h]
	\centering 
	\includegraphics[width=1\textwidth]{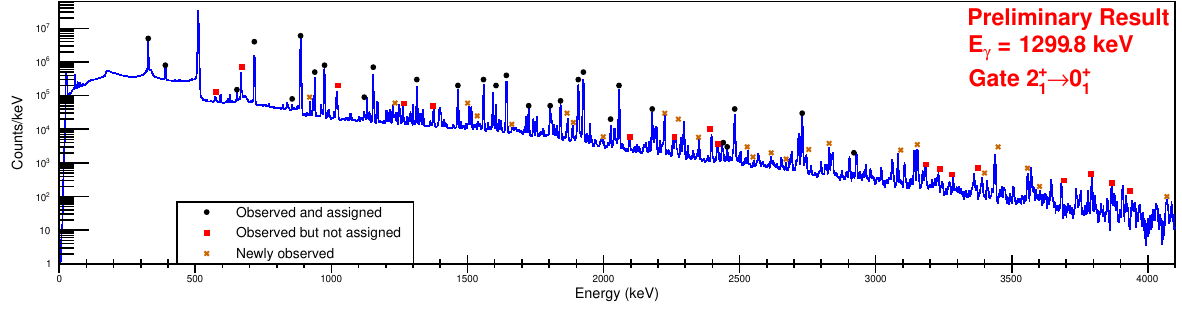}	
	\caption{Add-back spectrum of $\gamma$-rays in coincidence with the $2^+_1 \rightarrow 0_1^+$ 1299.8-keV ground state transition in $^{114}$Sn, showing selected transitions that were observed and assigned in literature (black circles), previously observed but not assigned in literature (red squares) 
\cite{PhysRevC.14.229, PhysRevC.97.054319, Araddad1990},  and newly observed (orange crosses) $\gamma$-ray transitions.} 
	\label{coincident}%
\end{figure*}

Figure \ref{coincident} shows a $\gamma$-ray spectrum in add-back mode, in coincidence with the 1299.8-keV transition, which corresponds to the decay from the first $2^+_1$ state to the ground state, $0^+_1$, of \ce{^{114}Sn}. In this context, coincidence refers to the detection of two $\gamma$ rays within a 250 ns time window, indicating that they are emitted from the same decay cascade. Many new transitions are observed, which must be incorporated into an expanded level scheme beyond 4 MeV.

\begin{figure}[h]
	\centering 
	\includegraphics[width=1\textwidth]{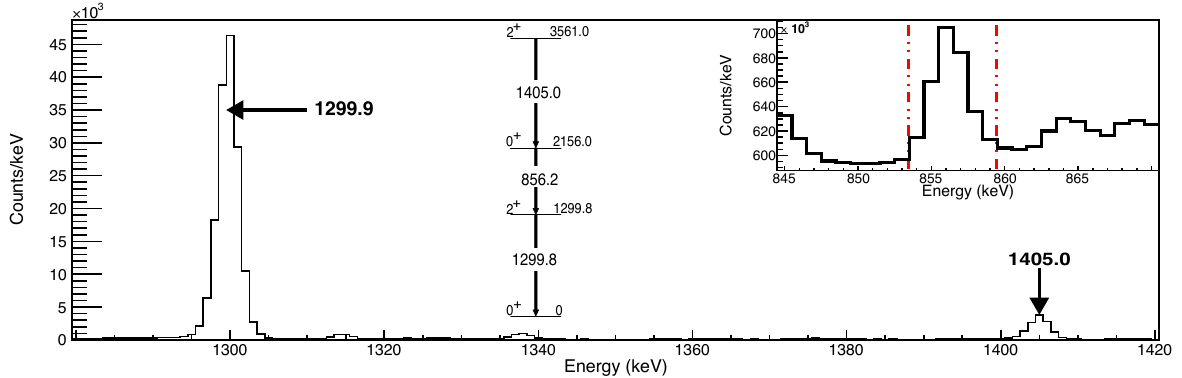}   	
	\caption{The $\gamma$-ray spectrum in coincidence with the 856.2-keV transition in $\ce{^{114}Sn}$. The inset shows the limits of the gate on the 856.2-keV transition. The presence of the 1299.8-keV and 1405.0-keV transitions allow for the building of the partial level scheme shown. This cascade shows the placement of the 1405.0-keV and 856.2-keV transitions, which had been observed in a previous $\beta^{+}$ decay experiment but had not been placed.} 
	\label{gate}%
\end{figure}

Figure \ref{gate} shows a portion of the $\gamma$-$\gamma$ coincidence spectrum, gated on 856.2-keV. This transition depopulates the $\ce{0_{3}^{+}}$ state at 2156.0-keV, and is shown to be in coincidence with the 1299.8-keV transition decaying to the ground state and a 1405.0-keV transition, which originates from 3561.0-keV, placed further above in the level scheme. Both the 856.2- and 1405.0-keV transitions were placed in the $^{114}${Sn}(n, n'$\gamma)$ experiment by Araddad \textit{et al.}~\cite{Araddad1990}, though they were not placed in the $\beta^{+}$ decay experiment by Wigmans \textit{et al.}~\cite{PhysRevC.14.229} despite being observed. The placement of these transitions, as observed in the $\beta^{+}$ decay using the GRIFFIN spectrometer, demonstrates the quality of the data available for investigations of the nuclear structure of $^{114}$Sn.

\section{Future Work}
% So far, for the $\beta$-decay of \ce{^{114}Sb} into excited states in \ce{^{114}Sn} significant work has gone into energy and efficiency calibration of the GRIFFIN spectrometer, including applying summing corrections and add-back, and preparing $\gamma -\gamma$ matrices for data analysis. The $\gamma -\gamma$ coincidences analysis of $^{114}$Sn data is still in progress, expanding the level scheme beyond 4 MeV.
Future work involves extracting $\gamma$-ray intensities to determine branching ratios and calculating log\textit{ft} values for both previously known and newly observed levels. These values, together with angular correlations, will aid in determining the spin states of newly observed energy levels. We will continue to verify existing transitions to ensure accuracy and search for new transitions to refine the level scheme. An article on the comprehensive spectroscopy of $\ce{^{114}Sn}$ including the decay patterns of the excited $0^+$ states will be published.

\section{Acknowledgements}
The infrastructure of GRIFFIN has been funded through
contributions from the Canada Foundation for Innovation,
TRIUMF, University of Guelph, British Columbia Knowledge
Development Fund, and the Ontario Ministry of Research and
Innovation. TRIUMF receives funding through a contribution
agreement through the National Research Council Canada.
This work was supported by the Natural Sciences and Engineering
Research Council of Canada. NKS acknowledges support from the Marie Sklodowska-Curie IAEA Fellowship Program.

\bibliographystyle{elsarticle-num}
\bibliography{ref}

\end{document}